\documentclass[conference]{IEEEtran}
		\IEEEoverridecommandlockouts
		\usepackage{cite}
		\usepackage{amsmath,amssymb,amsfonts}
		\usepackage{algorithmic}
		\usepackage{graphicx}
		\usepackage{textcomp}
		\usepackage{xcolor}
		\usepackage{times}
		\usepackage{changepage}
		\usepackage{color,soul}
		\usepackage{listings}
		\usepackage{color}
		\usepackage{fancyhdr}
		\usepackage{comment}
		\usepackage{wrapfig}
		\usepackage{multirow}\usepackage{multirow}
		\usepackage{multicol}\usepackage{multicol}
		\usepackage{subfig} 
		\usepackage{caption}
		\usepackage{morefloats}
		\usepackage{siunitx}
		\newcolumntype{C}[1]{>{\centering}m{#1}}
		\begin{document}
		
		\title{PHELP: Pixel Heating Experiment Learning Platform for Education and Research on IAI-based Smart Control Engineering\\
		}
		
		\author{\IEEEauthorblockN{Jairo Viola}
		\IEEEauthorblockA{\textit{MESA Lab} \\
		\textit{School of Engineering, UC Merced}\\
		Merced, California, USA \\
		jviola@ucmerced.edu}
		\and
		\IEEEauthorblockN{Carlos Rodriguez}
		\IEEEauthorblockA{\textit{MESA Lab} \\
			\textit{School of Engineering, UC Merced}\\
			Merced, California, USA \\
		crodriguezmartinez@ucmerced.edu}
		\and
		\IEEEauthorblockN{YangQuan Chen}
		\IEEEauthorblockA{\textit{MESA Lab} \\
			\textit{School of Engineering, UC Merced}\\
			Merced, California, USA \\
		ychen53@ucmerced.edu}}
		
		\maketitle
		
		\begin{abstract}
		
		Thermal processes are one of the most common systems in the industry, making its understanding a mandatory skill for control engineers. So, multiple efforts are focused on developing low-cost and portable experimental training rigs recreating the thermal process dynamics and controls, usually limited to SISO or low order 2x2 MIMO systems. This paper presents PHELP, a low-cost, portable, and high order MIMO educational platform for uniformity temperature control training. The platform is composed of an array of 16 Peltier modules as heating elements, with a lower heating and cooling times, resulting in a 16x16 high order MIMO system. A low-cost real-time infrared thermal camera is employed as a temperature feedback sensor instead of a standard thermal sensor, ideal for high order MIMO system sensing and temperature distribution tracking. The control algorithm is developed in Matlab/Simulink and employs an Arduino board in hardware in the loop configuration to apply the control action to each Peltier module in the array. A temperature control experiment is performed, showing that the platform is suitable for teaching and training experiences not only in the classroom but also for engineers in the industry. Furthermore, various abnormal conditions can be introduced so that smart control engineering features can be tested.
				
		\end{abstract}
		
		
		\section{Introduction}
		Performing a precise temperature control is one of the main objectives in the industry due to its relevance in many manufacturing processes like oil refining \cite{c1}, food production \cite{c2}, agriculture \cite{c3}, semiconductors manufacturing \cite{c4} among others. Thus, its study and analysis are a mandatory subject during the education process of control engineers.    So,  part of the control research is focused on developing portable temperature processes as a training platform representing real industrial temperature control systems.
		\par
		For this reason, there are different training rigs for temperature control that can be employed not only in academia but also for industry \cite{c5}-\cite{c96}. However, many of these platforms are expensive, require laboratory arrangements, and have higher heating and cooling respond times, making it not efficient for teaching and training requiring several hours for a session to be completed.   These training platforms are not open hardware and software, limiting the possibility of further implementing advanced control strategies than PI or PID, as well as incorporating additional sensors or control elements for further research and behavior assessment. 
		Most of the current temperature training platforms are limited to SISO control loops or 2x2 MIMO systems. It means that there is no possibility of working with high order MIMO systems with coupling and interaction among its control loops.
		\par
		This paper present the Pixel Heating Experiment Learning Platform PHELP, a low-cost real-time high order MIMO temperature uniformity control platform for education and research, developed on the Mechatronics, Automation and Embed systems laboratory (MESALab) of the University of California, Merced. PHELP is composed of a set of 16 Peltier modules controlled individually in a planar array configuration, which works as heating and cooling elements. A low-cost infrared thermal camera is employed as a feedback sensor for the system, which gives a visual distribution of the temperature uniformity on the system. The control system is implemented in Matlab-Simulink and is configured in hardware in the loop (HIL) with an Arduino board to send the control actions to the system. Likewise, the platform is open for the implementation of different control strategies and the integration of more sensing devices and actuators. 
		\par
		The main contribution of this paper is developing for the first time of a low-cost real-time High-order uniformity temperature control platform with an array of Peltier as heating elements designed for training and teaching on Multi-Input Multi-Output (MIMO) systems, with the possibility of the implementation of different control strategies as well as it is easy transportation and installation. Likewise, defining a methodology to solve a temperature control problem, following the steps of data acquisition, system identification, controller design, and implementation. Thus, the control engineering students will have the tools to take and solve a MIMO temperature control problem.
		\par
		The paper is structured as follows. Initially, the PHELP system is presented. Then, a MIMO temperature control experiment is performed to show the main features of PHELP. Finally, conclusions and future works are presented.
		
		\section{PHELP system description}
		
	    The PHELP training platform is presented in Fig.1 and its block diagram in Fig. 2. As can be observed, the system is composed by an array 4 by 4 of Peltier modules (P5), a thermal infrared camera operating on a Raspberry Pi (P4) which send the temperature data using TCP/IP protocol, and an Arduino board (P1) configured on hardware in the loop (HIL) configuration with Matlab-Simulink to perform the data acquisition, identification and control tasks for the system. A power block is included formed by a PWM generator and a power driver (P2) to manage the power applied to each Peltier module according to the control action defined by the control algorithm. Between the Peltier module and the Power Driver, a module of power resistors (P3) is used to dissipate the power in the line and guarantee the maximum voltage and current handle by the Peltier. A detailed description of the PHELP system components is presented in the following sections.
				\begin{figure}
				\centering
				\includegraphics[width=0.28\textwidth,height=0.25\textheight]{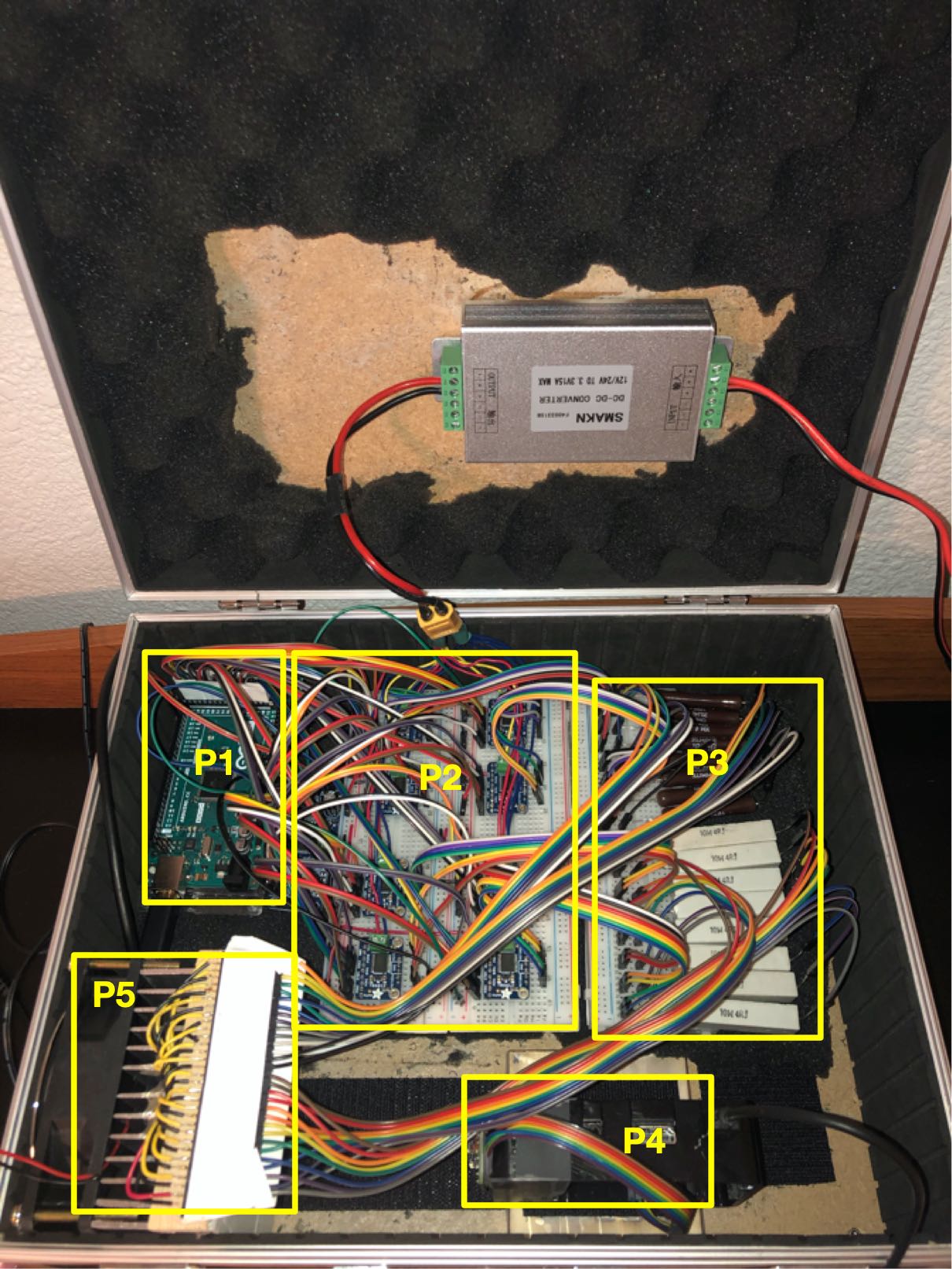}
				\caption[caption]{PHELP Experimental Platform}
			\end{figure}
			\begin{figure}
				\centering
				{\includegraphics[width=0.45\textwidth,height=0.25\textheight]{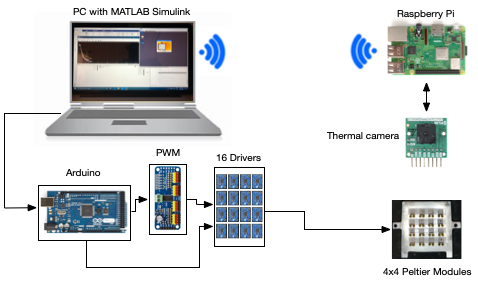}}\\
				\caption[caption]{PHELP platform block diagram}
			\end{figure}
		
			\subsection{Peltier module array}\label{AA}
\begin{figure}
				\centering
				\includegraphics[scale=0.03]{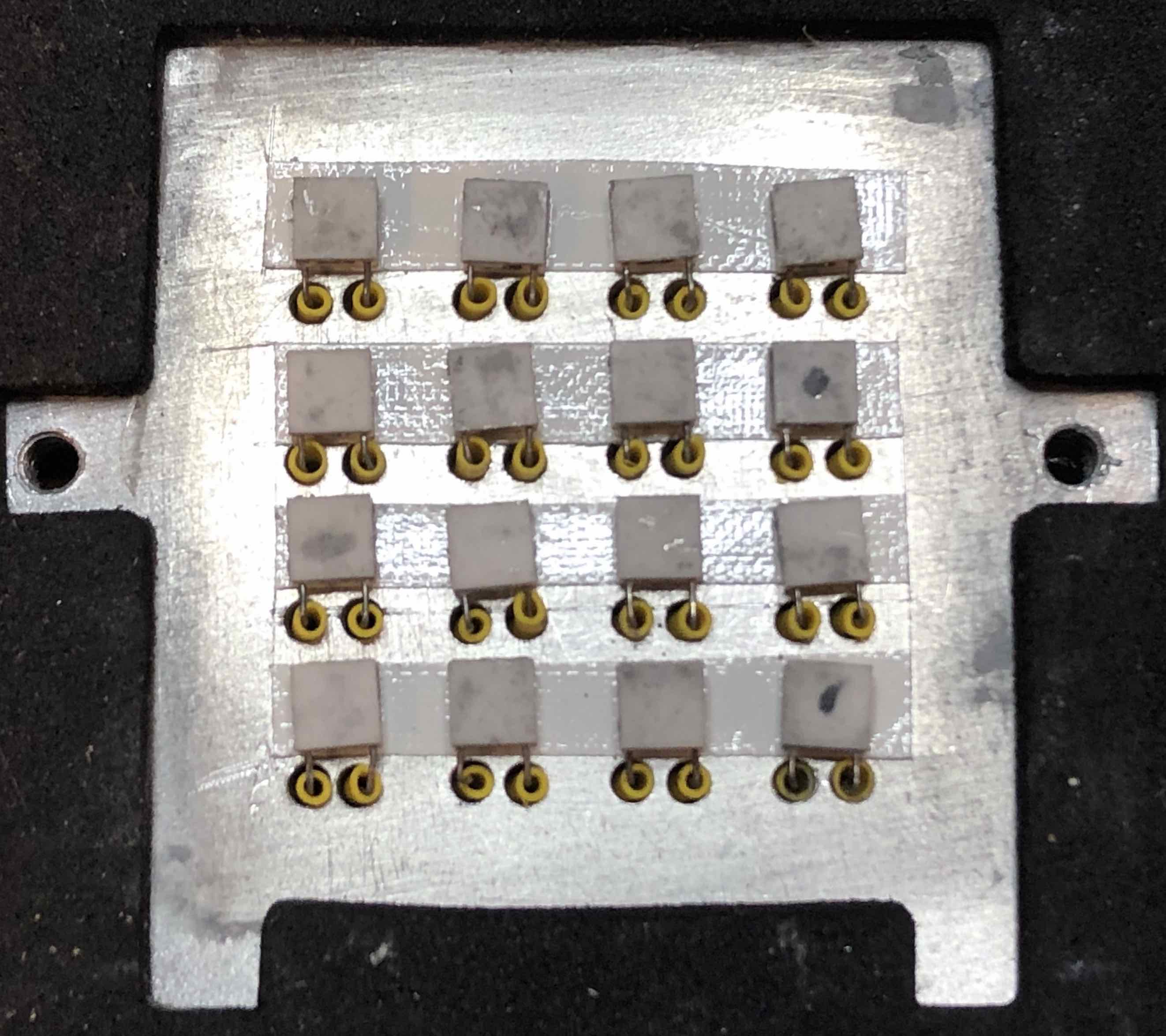}
				\caption[caption]{PELTIER MODULE ARRAY}
				\label{array}
			\end{figure}

		    The 16 Peltier array employed in the PHELP system is presented in Fig. \ref{array}. As can be observed, each Peltier thermoelectric cooler (TEC)  is a solid-state heat pump that transfers heat from one side of the device to another depending on the direction of the current. When current is applied, one side gets hot, and the other gets cool. It used to be employed for cooling, heating, or temperature regulation. \par 
			The Peltier array was constructed with the NL1020T module employed as heating elements because it is a solid-state device with low maintenance requirements and long service lifetime.
			The temperature range for this device comes from $15^oC$ to $100^oC$, with maximum heating of $1W$ and a power requirement of $0.9 VDC$ $1.8 A$. For the MESALab temperature platform, the Peltier system's power is controlled using pulse width modulation managed by the Arduino board, PWM/Servo Driver, and power driver. PWM signal range goes from $-4000$ to $0$ for cooling, and from $0$ to $4000$ for heating. Over the Peltiers shown in Figure \ref{array}, an aluminum surface is installed to create the interaction between the 16 Peltiers. The temperature control system regulates the value in the place of each Peltier depend on the desired value; always be physically realizable. A heat sink was modified to install all the Peltiers and dissipate the extra heat to extend the modules lifetime. \par
			This design generates a dynamic MIMO system with 16 inputs and outputs, which is a complex system due to the interaction between the Peltier modules, which can be analyzed like a perturbation to each input$_i$-output$_i$ relation.
		
		\subsection{Infrared thermal camera}
\begin{figure}
			\centering
			\includegraphics[scale=0.25]{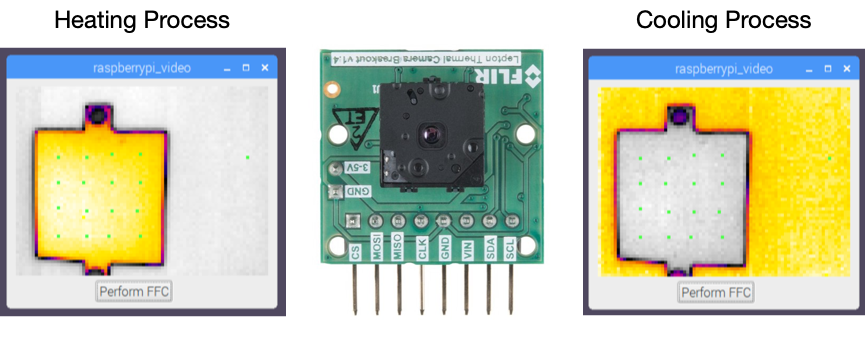}
			\caption[caption]{FLIR Lepton thermal camera}
			\label{camera}
		\end{figure}

        Figure \ref{camera} shows the thermal infrared camera employed as a temperature sensor for this system, to the left and image of the heating process, and the right of the cooling process, in both images,  green dots were set to represent the measurements points of the system, which is aligned with the Peltiers modules. The camera manufactured by FLIR is a long-wave infrared camera that measures the temperature over a surface through its infrared emitted radiation \cite{c10}. This camera's wavelength range comes from $8\mu m$ to $14 \mu m$ with a maximum frame rate of 9 FPS. The camera has a resolution of 80x60 pixels with an accuracy of $\pm0.5^oC$, and its size is less than a quarter coin. Besides, the camera has an SPI interface, which allows its connection with many edge devices. Also, the LeptonThread software development kit is available for the camera data acquisition, which runs in python and C++. For this platform, the thermal camera works together with a Raspberry Pi 3B+, which reads the camera through the SPI interface, sending the data to Matlab-Simulink employing a TCP/IP client-server configuration. In this system, the thermal camera with the Raspberry Pi acts as the server, and Matlab-Simulink application runs as the client for the thermal data camera. An example of the infrared vision camera image working on the Raspberry Pi is presented in Figure \ref{camera}.
		
		\subsection{PHELP power management interface}
\begin{figure}
			\centering
			\includegraphics[scale=0.1]{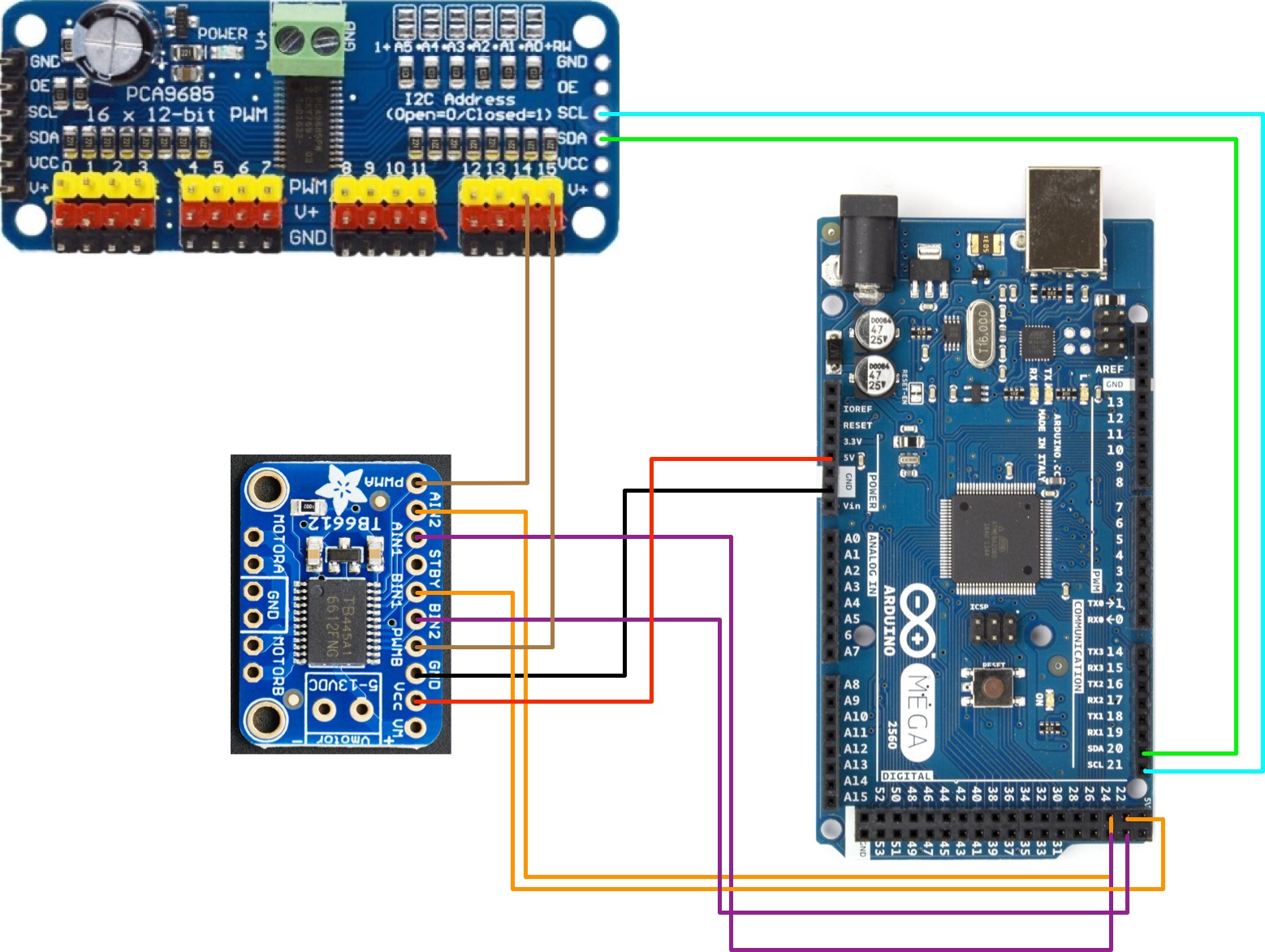}
			\caption[caption]{Arduino Mega, PWM generator and TB6625 motor driver}
			\label{devices}
		\end{figure}

	    Figure \ref{devices} shows the power management interface employed in the PHELP system. It is composed of an Arduino Mega board working as a data acquisition interface with  MATLAB\textregistered-Simulink software. It is configured as HIL in order to obtain a real-time interaction of the real system with the PC. After sending the control signal through the Arduino, this signal split in magnitude (to the PWM/Servo Driver) and direction (to power driver). Finally, the PWM/Servo Driver applies the PWM signal to the power driver to control the current in each Peltier and the temperature as a consequence. The Adafruit 16-Channel 12-bit PWM/Servo Driver \cite{c13} was used to communicate with the Arduino board using I2C protocol, and the Adafruit TB6612 1.2A DC/Stepper Motor Driver with a range of $4.5V DC$ to $13.5V DC$ \cite{c14}.
		
		\section{Matlab Simulink interface}
		PHELP employs Matlab-Simulink for data acquisition, identification, and control tasks performed during the temperature control exercises.  The purpose of employing Matlab-Simulink for this platform is enabling the capability of implementing different control algorithms from the classic PI and PID controllers to robust, optimal, and nonlinear control strategies, using classic tools well known by most of the control engineers on academia or industry without the need of embedded implementation.    \par 
		On the other hand, Matlab-Simulink allows interaction with different sensors and control boards with different communication protocols. In this particular case, the infrared camera uses the TCP/IP communication protocol, and the Arduino board uses a serial communication protocol, both of them supported by Matlab-Simulink. So, adding new sensors and instruments to the system is possible without further complexity.	
		
		\subsection{Total cost of the system}\label{AB}
	    Table 1 shows the total cost of this platform. As can be observed, the platform's overall cost is $727.00 \$$ USD, which is an affordable cost for many engineering schools interested in MIMO complex temperature control training.
		\begin{table}
			\centering 
			\caption{Temperature training platform total cost}
				\label{table_ASME}
				\begin{tabular}{c c c}
					\hline
					Component & Quantity & Total Price \\ \hline
					16 Peltier module & 16 & \$288.00\\
					Raspberry PI 3B+ & 1  & \$35.00 \\
					Arduino Mega board& 1 & \$30.00\\
					PWM/Servo Driver& 1 & \$15.00 \\
					Dual Power drivers& 8 & \$80.00 \\
					Thermal Camera & 1 & \$229.00 \\
					Others &  & \$40.00 \\ \hline
					Total cost & & \$727.00
				\end{tabular}
		\end{table}
		
		\section{PHELP temperature control training experiment}
		The training experiment proposed in this paper tests the PHELP platform's capabilities for designing and implementing a MIMO temperature control system. The experiment involves the stages of multivariable system identification, PID controller design using FOMCON Toolbox \cite{c15} base in the methodology of \cite{c16}, and its practical implementation using the HIL architecture.
		
		\subsection{Multivariable system identification}\label{AC}
\begin{figure}[h]
			\centering
			\includegraphics[scale=0.3]{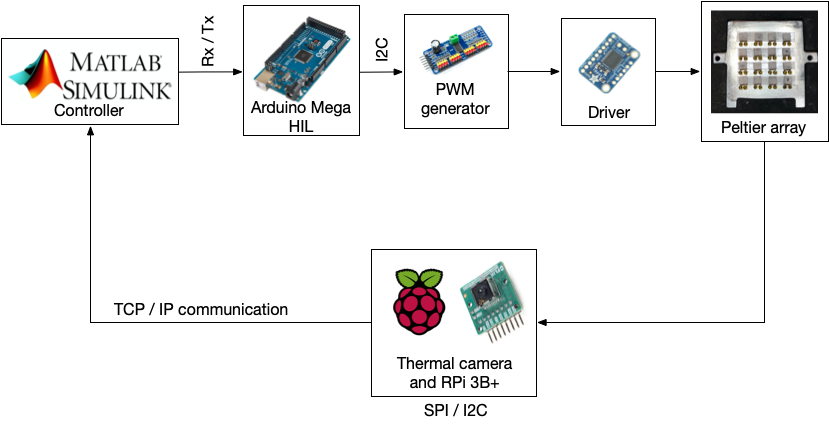}
			\caption[caption]{Temperature system closed loop configuration}
			\label{closedloop}
		\end{figure}

\begin{figure}
			\centering
			\includegraphics[width=6cm]{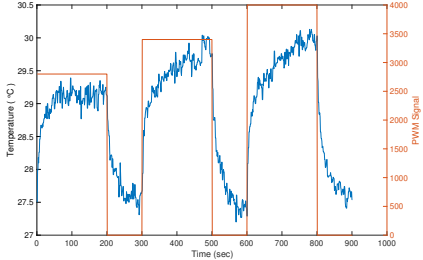}
			\caption[caption]{Identification signal for the peltier module}
			\label{steppeltier}
		\end{figure}

	    Figure \ref{closedloop} shows the block diagram of the system configuration employed for the temperature training experiment. The Peltier array module is the plant to be controlled. It is characterized by a MIMO model where an integer or fractional-order system models the interaction between the Peltier through the aluminum surface. The thermal infrared camera is the temperature feedback sensor. 
		A robust identification experiment is designed for the MESALab temperature system in order to identify its uncertainty. This experiment consists of applying stepped signals with different amplitudes. The dynamic behavior of each Peltier and the interaction between them was measurement in the 16 Peltiers for a total of 256 data vectors. The validation of the system model in Figure \ref{validation} is performed by applying a step signal to all the 16 Peltiers simultaneously. Figure \ref{steppeltier} shows the identification data acquired from the Peltier for the first row and column of the Peltier array shown in Fig. \ref{array}. \\
        The system is identified using a transfer function array defined by \eqref{model} with the structure of a Fractional Order Plus Dead Time system FOPDT is given by \eqref{tfsystem}, where $K$ is the system gain, $T$ is the system time constant, $\alpha$ is the fractional order of the denominator, and $L$ is the time delay for the system for the $i$ and $j$ columns of the transfer function array.\par
\begin{figure}
			\centering
			\includegraphics[scale=0.3]{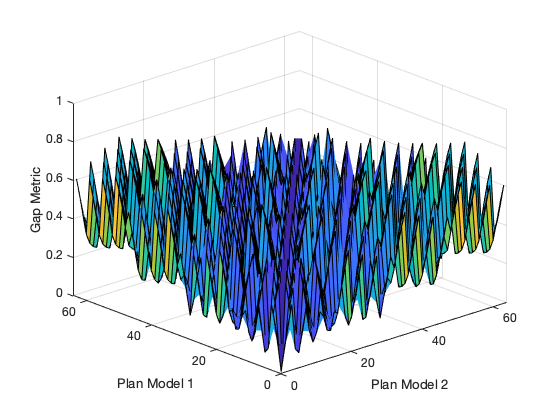}
			\caption[caption]{Gap Metric surface}
			\label{gap}
		\end{figure}
\begin{figure}
			\centering
			\includegraphics[scale=0.3]{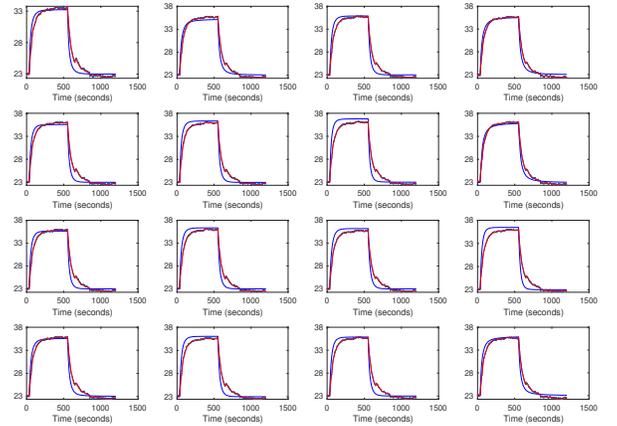}
			\caption[caption]{Validation of the PHELP system model}
			\label{validation}
		\end{figure}

		Equations \ref{modeluncertain} show that the system presents different models for each stepped signal applied. Thus, it is possible to define the family of plant for the temperature system with the uncertainty boundaries for $K, T$ and $\alpha$, defined by \eqref{tfsystem}. 
		\begin{equation}
		\begin{split}
			G_1(s)=\frac{1.3026}{15.153s^{0.9}+1} \\
			G_2(s)=\frac{2.3329}{8.3473s^{0.6}+1} \\
			G_3(s)=\frac{3.3999}{6.7947s^{0.5}+1} 
		\end{split}
		\label{modeluncertain}
		\end{equation}
		After that, the gap metric among all the plants is calculated, producing a gap metric matrix, using the Oustaloup approximation of fractional order systems for CRONE. The gap metric surface for the set of plants is shown in Figure \ref{gap}, where its possible to see that the gap metric converges close to zero. For the first Peltier, the nominal model is given by \eqref{nominal}, and the rest of the model appears in Table \ref{table2}.
		\begin{equation}
			G(s)=\frac{1.3026}{12.367s^{0.5}+1}
			\label{nominal}
		\end{equation}
	
		Considering that the system has 256 transfer functions, Table \ref{table2} presents the model parameters for the main diagonal of $Y(s)$. For these cases, the term $L$ is equal to zero because of the system in the present diagonal delay. All the PHELP system identification data and parameters are available in \cite{c12}.\par
		The validation of the MIMO model \eqref{model} is shown in Figure \ref{validation}, where the blue lines are the model outputs, and the red lines are the real validation data. As can be observed, the identified model has fitness with the validation data.	
		\begin{equation}
		Y(s)=\left[\begin{matrix}
		G_{1,1} (s)& G_{1,2}(s) & \dots & G_{1,16}(s)\\
		G_{2,1} (s)& G_{2,2}(s) & \dots & G_{2,16}(s) \\
		\vdots(s)& \vdots & \ddots & \vdots \\
		G_{16,1} (s)& G_{16,2}(s) & \dots & G_{16,16}(s) \\
		\end{matrix}\right]U(s)
		\label{model}
		\end{equation}
		
		\begin{eqnarray}
		G_{ij}(s)=\frac{K_{ij}}{(T_{ij}s^{\alpha_{ij}}+1)}e^{-L_{ij}s},~~~~~i,j=1,2,\dots,16.
		\label{tfsystem}
		\end{eqnarray}
		
		\begin{table}
			\centering
			\caption{$Y(s)$ main diagonal transfer functions parameters}
			\label{table2}
			\begin{tabular}{|c|C{0.5cm}|C{0.5cm}|C{0.65cm}|C{0.5cm}|C{0.65cm}|C{0.65cm}|C{0.5cm}|c|}
				\hline
				&$G_1$&$G_2$&$G_3$&$G_4$&$G_5$&$G_6$&$G_7$&$G_8$ \\ \hline
				K& 1.30 & 1.61&0.79&1.83&1.45&0.67&0.26&0.72\\ \hline
				T& 12.36&	20.13&	17.89&	20.82&	21.51&	35.43&	9.10&	11.60\\ \hline 
				$\alpha$& 	0.5&	0.75	&0.86	&0.5	&0.65	&0.85	&0.65	&0.65	 \\ \hline
				&$G_9$&$G_{10}$&$G_{11}$&$G_{12}$&$G_{13}$&$G_{14}$&$G_{15}$&$G_{16}$ \\ \hline
				K&	 1.11&1.41&0.92&0.61&0.94&0.69&0.30&0.73\\ \hline
				T&	 23.12&	18.38&	28.27&	16.58&	8.41&	15.62&	7.57&	5.34\\ \hline
				$\alpha$& 	0.61&	0.5&	1&	0.9&	0.65&	0.5&	0.5&	0.9\\ \hline
			\end{tabular}
		\end{table}
		
		\begin{table}
			\centering
			\caption{Model FIT}
			\label{fit}
			\begin{tabular}{|c|C{0.6cm}|C{0.6cm}|C{0.6cm}|C{0.6cm}|C{0.6cm}|C{0.6cm}|c|}
				\hline
				$M_1$&$M_2$&$M_3$&$M_4$&$M_5$&$M_6$&$M_7$&$M_8$ \\ \hline
				76.03 & 79.88 & 80.61 & 83.43 & 78.02 & 78.00 & 73.93 & 86.59  \\ \hline
				$M_9$&$M_{10}$&$M_{11}$&$M_{12}$&$M_{13}$&$M_{14}$&$M_{15}$&$M_{16}$ \\ \hline
				75.98 & 78.39 & 79.96 & 75.33 & 78.99 & 76.83 & 76.08 & 80.87 \\ \hline 
			\end{tabular}
		\end{table}
		
		\subsection{PHELP system controllers design}
	    In this paper, a decentralized PI control strategy controls the temperature on the PHELP system. Thus, the FOPDT models of the main diagonal of the transfer function matrix \eqref{model} are employed for controller design using the parameters in Table \ref{table2}. An optimal PI tuning algorithm \cite{c16} is employed to find the PI controller gains. It requires an LTI integer or fractional order transfer function of the system $G_p (s)$ and the controller transfer function $G_c (s)$, which must satisfy the following specifications:
		\begin{itemize}
			\item Phase margin (pm):
			\begin{equation}
			arctan(G_c(jw)G_p(jw))=-\pi+pm
			\label{spec1}
			\end{equation}
			
			\item Gain crossover frequency($w_c$):
			\begin{equation}
			|G_c(s)G_p(s)|=0~dB
			\label{spec2}
			\end{equation}	
			
			\item Robustness against variation of plant gain:
			\begin{equation}
			\frac{d}{dw}arctan(G_c(jw)G_p(jw))=0
			\label{spec3}
			\end{equation}
			
			\item Rejection of high frequency noise:
			\begin{equation}
			\left | \frac{G_c(jw)G_p(jw)}{1+G_c(jw)G_p(jw)} \right |=B~dB
			\label{spec4}
			\end{equation}
			
			\item Rejection of output disturbances:
			\begin{equation}
			\left | \frac{1}{1+G_c(jw)G_p(jw)} \right |=A~dB
			\label{spec5}
			\end{equation}
			
		\end{itemize}	
        The Matlab FOMCON Toolbox is employed to solve the optimal control problem to find the values of $K$ and $T_i$ using the PI controller form \eqref{pidequation}. Thus, the desired system performance specifications are given in terms of gain margin and phase margin using the Integral of Time multiplied by Absolute Error (ITAE) criterion as the optimization criterion. For these models, a phase margin $ PM = 60^\circ - 65^\circ,$ and a gain margin $GM = 10 - 15\, dB.$ are employed, and the obtained parameters of the PI controllers are shown in Table \ref{fit}.
		\begin{equation}
		PI(s)=K\left(1+\frac{1}{T_is}\right)=K+\frac{I}{s}.
		\label{pidequation}
		\end{equation}
		
		\begin{table}
			\centering
			\caption{Decentralized PI controller parameters}
			\label{parameters}
			\begin{tabular}{|c|C{0.5cm}|C{0.5cm}|C{0.65cm}|C{0.5cm}|C{0.65cm}|C{0.65cm}|C{0.5cm}|c|}
				\hline
				&$PI_1$&$PI_2$&$PI_3$&$PI_4$&$PI_5$&$PI_6$&$PI_7$&$PI_8$ \\ \hline
				K & 65.73 & 83.33 & 83.33 & 83.33 & 83.33 & 83.33 & 83.33 & 83.33  \\ \hline
				I & 1.17 & 2.01 & 2.44 & 2.83 & 1.5 &  1.5&1.5   &1.5  \\ \hline 
				&$PI_9$&$PI_{10}$&$PI_{11}$&$PI_{12}$&$PI_{13}$&$PI_{14}$&$PI_{15}$&$PI_{16}$ \\ \hline
				K & 83.33 & 68.57 & 83.33 & 83.33 & 83.33 & 83.33 & 83.33 & 83.33 \\ \hline
				I &  1.5&  1.23&   3.34  &  2.35  &  1.5&    1.5&  1.5  &  4.45\\ \hline
			\end{tabular}
		\end{table}
		
\begin{figure}[ht]
			\centering
			\includegraphics[scale=0.2]{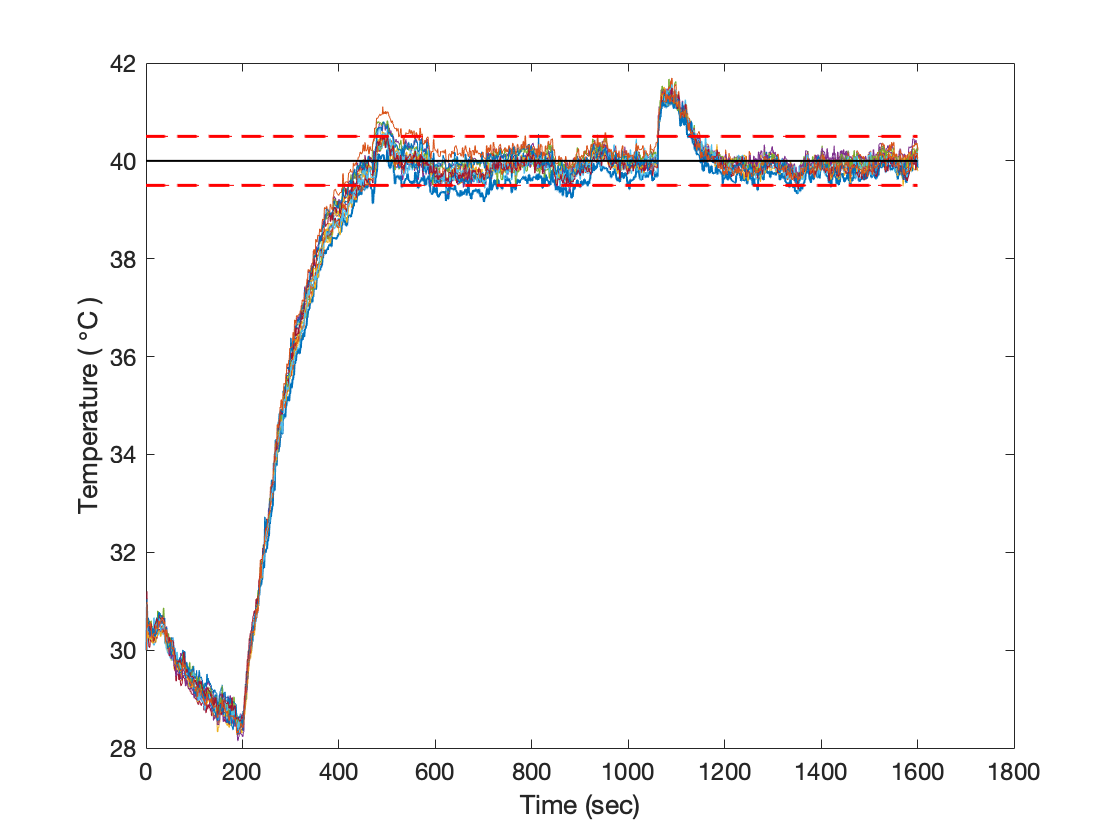}
			\caption[caption]{PHELP system step response with PI controllers}
			\label{pidtemp}
		\end{figure}

\begin{figure}[ht]
			\centering
			\includegraphics[scale=0.2]{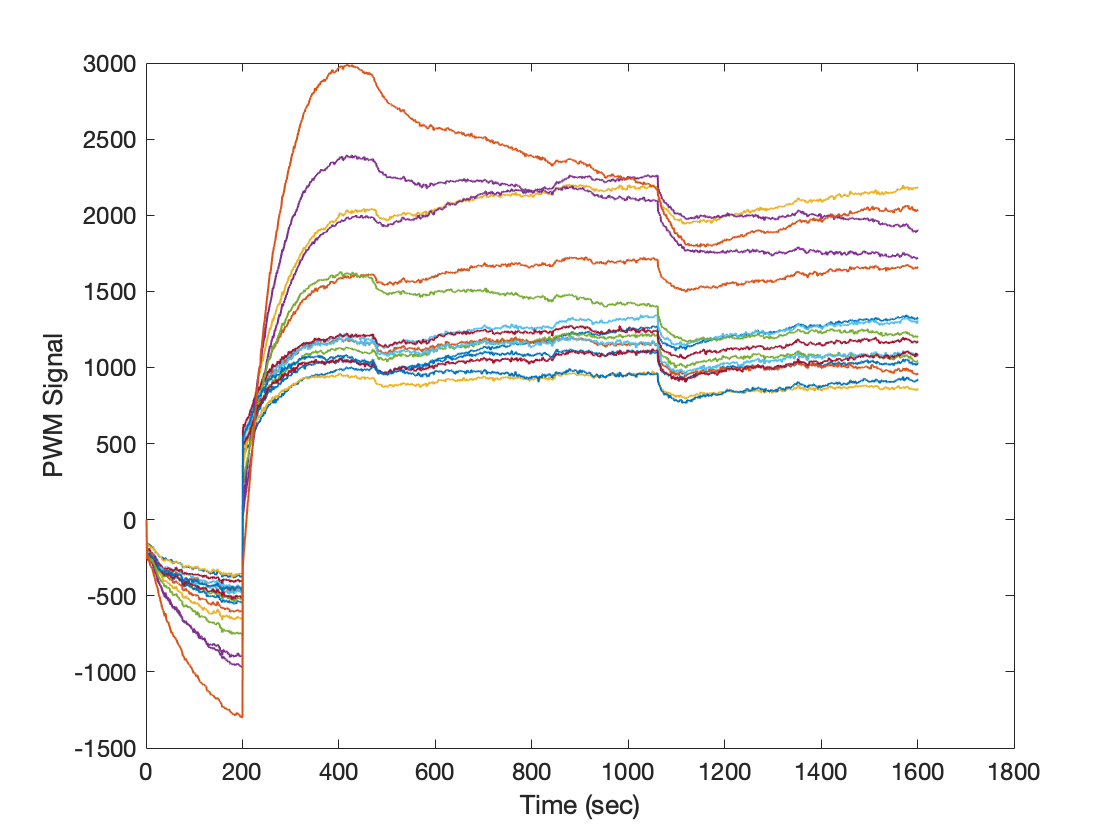}
			\caption[caption]{PHELP system control action for PI controllers}
			\label{pidcontrol}
		\end{figure}

		Figure \ref{pidtemp} and Fig. \ref{pidcontrol} shows the time response and control action of the system for a change in the reference around ($13\, ^\circ$C). As can be observed, the decentralized PI controllers reach the desired setpoint with uniformity in the control signal, allowing a suitable thermal performance in the surface to avoid wearing the material and prolong the life cycle of the Peltier. At the time of 1061 seconds, the thermal camera applies a perturbation in the control system due to the thermal camera presents an error of $\pm 0{.}5\, ^\circ$C, and every certain time it performs an adjustment of the image quality, changing the measured value markedly.
		
		\section{PHELP as a Smart Control Engineering testbed}
\begin{figure}[htb]
			\centering
			\includegraphics[scale=0.12]{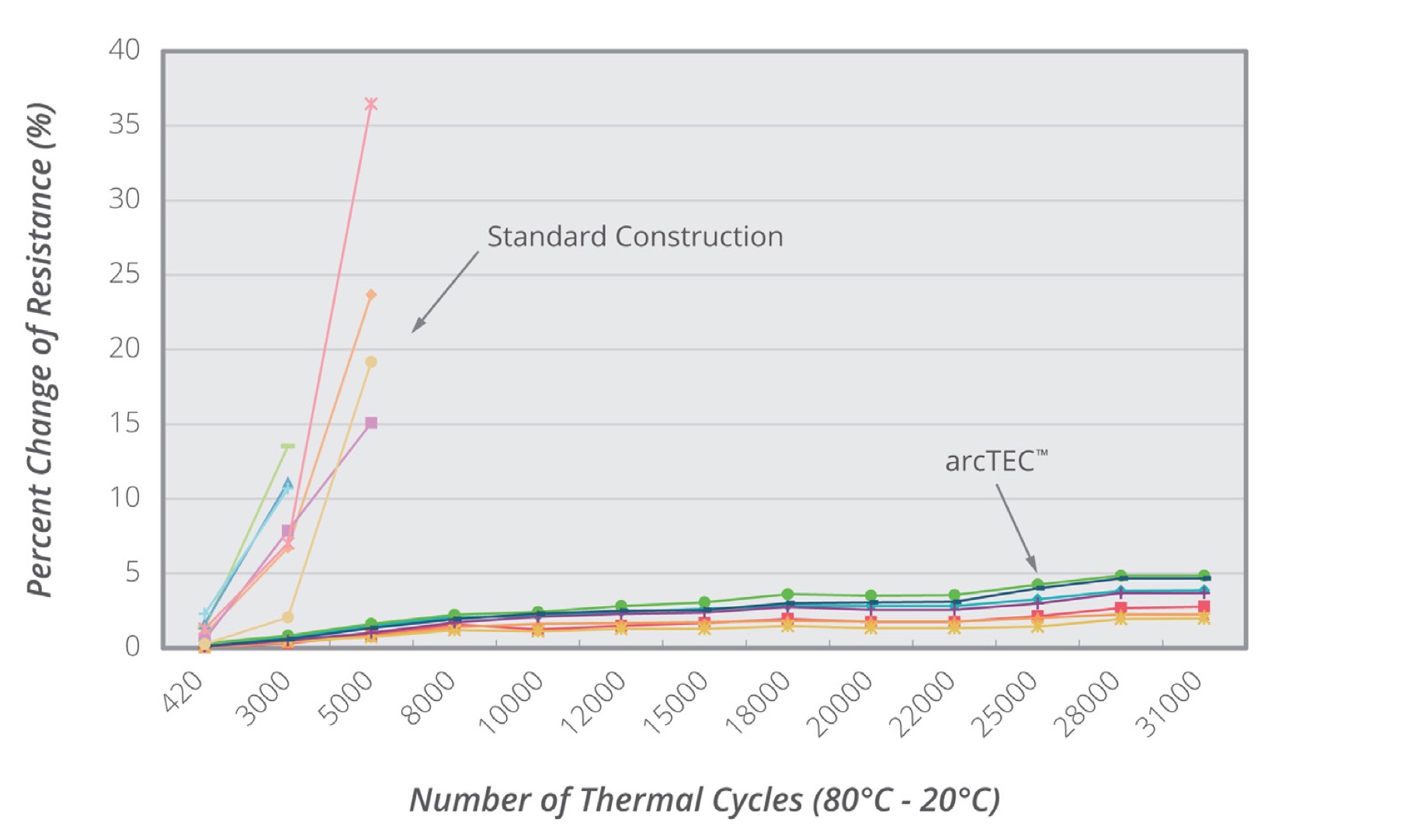}
			\caption[caption]{Reliability test}
			\label{lifecycle}
		\end{figure}

		Digitalization and new technologies, such as digital twins and artificial intelligence, allow industrial companies to develop speed, efficiency, quality, and flexibility in an unprecedented way. Ideally, the digital design tools integrate into the real-world control of the production facility or the product or prototype in question. It allows testing different production scenarios and validating the changes before taking the new features into production. \\
		Besides, according to \cite{c17}, a smart control system has the following characteristics: Cognizant, Taskable, Reflexive, Knowledge Rich, and Ethical. The system is considered cognizant if it is aware of its capabilities and limitations to face changes and variability in the system and its environment. Likewise, the system is taskable if it can process high-level natural language commands produced by an automated stimulus or human commands.  Also, the system is Reflective if it can learn from previous experiences to improve its current and future performance. Besides, a system is knowledge-rich when it can be reasoning based on multiple distributed and diverse sensor information. Finally, a system is ethical if its behavior is attached to the social and legal norms. \\
		In the case of the PHELP platform, its features open the system to be an IAI and smart control engineering dataset. For instance, the use of edge computing devices like Raspberry Pi and Lattepanda allows introducing artificial intelligence in the loop, from the infrared real-time feedback camera, performing thermal imaging smart recognition of temperature and failures. These failures are linked to the thermal fatigue of the Peltier device that affects its reliability \cite{c18} Figure \ref{lifecycle} and by interruptions in the Peltier power supply. 
		 \\
		Considering the size of PHELP (16x16 MIMO), the system becomes a rich source of information for monitoring multiple sections of the system, introducing current, voltage, among other sensors useful to build big data from the system.
		\\
		In addition, considering that the PHELP's main objective is the uniformity temperature control, there is a high-level task that can be given to the system combined with AI processors that set the desired heat pattern of the system.
		\\
		On the other hand, the Infrared thermal camera information allows modeling the system as a diffusion equation, transforming into an infinite-dimensional system, opening the system for advanced modeling techniques and representations like Digital Twin.

		\section{Conclusions}
		The low-cost MIMO high order PHELP platform for temperature control was introduced for training and learning experiences. The system is a 16x16 MIMO system that uses 16 Peltier modules as heating elements. The control is performed using Matlab with Arduino in hardware in the loop configuration. The system's overall cost is $727.00\$ USD $, making it affordable for many engineering schools. The experimental results show that the system has a MIMO behavior, that offers the control engineers a reference system applying modeling, analysis, design, and implementation methodology for control system design and implementation. Therefore, based on the obtained results, we can say that PHELP is a suitable platform for MIMO coupled training and learning experiences in academia and industry. As future works, the development of a Digital Twin for remote teaching as well as networked interaction for the PHELP system is proposed.


		\end{document}